\documentclass{iopart}
\usepackage{graphicx}
\usepackage{amssymb}

\begin{document}
\title{The Quantum Interference Effect Transistor: Principles and Perspectives}
\author{Charles A. Stafford$^1$, David M. Cardamone$^2$, and Sumit Mazumdar$^1$}

\address{$^1$ Department of Physics, University of Arizona, 1118 E. 4th
  Street, Tucson, Arizona, United States, 85721}
\address{$^2$ Department of Physics, Simon Fraser University, 8888 University Drive,
Burnaby, British Columbia, Canada, V5A 1S6.}
\ead{stafford@physics.arizona.edu}

\begin{abstract}
We give a detailed discussion of the Quantum Interference Effect Transistor (QuIET), 
a proposed device which exploits interference between electron paths through aromatic
molecules to modulate current flow.  In the off state, perfect destructive interference
stemming from the molecular symmetry blocks current, while in the on state, current
is allowed to flow by locally introducing either decoherence or elastic scattering.
Details of a model calculation demonstrating the efficacy of the QuIET are presented,
and various fabrication scenarios are proposed, including the possibility of using
conducting polymers to connect the QuIET with multiple leads.
\end{abstract}

\pacs{
85.65.+h, 
73.63.-b, 
31.15.Ne, 
03.65.Yz  
}

\submitto{\NT (NGC2007 Special Issue)}

\maketitle

\section{Introduction}
Despite their low cost and extreme versatility, modern semiconductor
transistors face fundamental obstacles to continued
miniaturization. First, top-down fabrication gives them microscopic
variability from device to device, which, while acceptable at today's
length scales, renders them unscalable in the nanometre regime. Second,
these devices, like all field effect devices, function by raising and lowering
an energy barrier to charge transport of at least $k_BT$; each device
therefore dissipates energy of this magnitude into the environment with every
switching cycle. At device
densities greater than the current state of the art, the cost and
engineering challenges associated with removing the resultant heat are
daunting \cite{roadmap}. While the first challenge can be met by utilizing the bottom-up,
chemical fabrication of single-molecule devices, this approach in itself does
nothing to address the need for a cooler switching mechanism.

An alternative paradigm to raising and lowering an energy barrier is to
exploit the wave nature of the electron to control current
flow \cite{sautet88,sols89,c60,baer02,ratner02,stadler03}. Traditionally, such interference-based devices are modulated via the
Aharanov--Bohm effect \cite{washburn86}. This, however, is incompatible with the small size of
molecular devices \cite{sautet88}: Through a nm$^2$ device, a magnetic field of over 600T
would be required to generate a phase shift of order 1 radian. Similarly, a
device based on an electrostatic phase shift \cite{baer02} would require voltages
incompatible with structural stability. Previously \cite{QuIETNL}, we have proposed a
solution, called the quantum interference effect transistor (QuIET) (see
Figure 1), which
exploits a perfect destructive interference due to molecular symmetry and
controls quantum transport by introducing decoherence or elastic scattering.

\begin{figure}
  \includegraphics[keepaspectratio=true,width=\columnwidth]{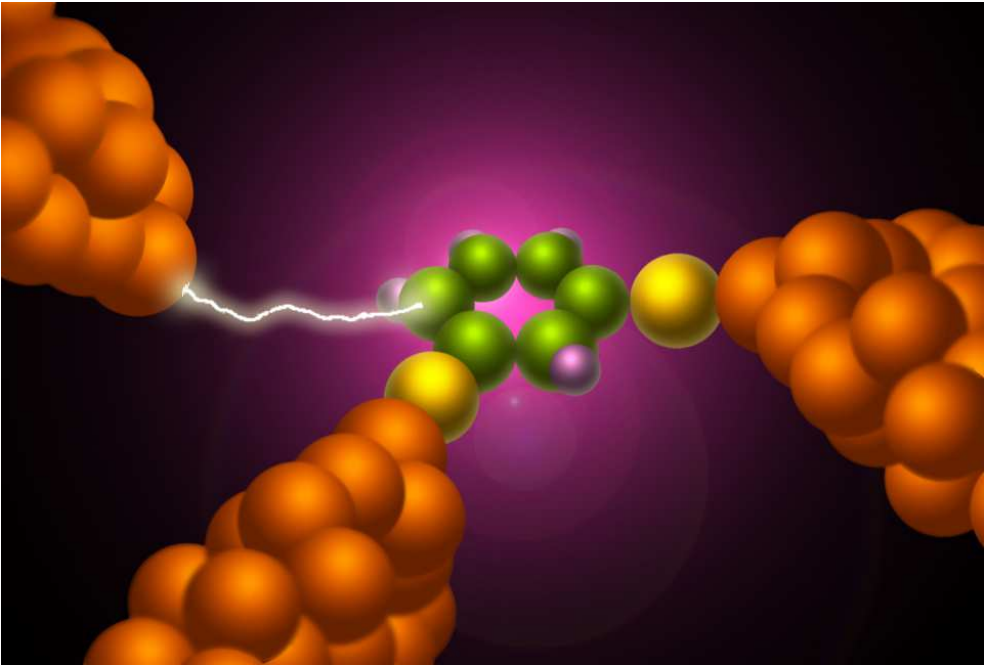}
  \caption{Artist's conception of a quantum interference effect transistor
    based on 1,3-benzenedithiol. The coloured spheres represent
    individual carbon (green), hydrogen (purple), sulfur (yellow), and gold
    (gold) atoms.  In the ``off'' state of the device, destructive interference blocks the
    flow of current between the source (bottom) and drain (right) electrodes.
    Decoherence introduced by the STM tip (upper left) suppresses interference,
    allowing current flow.  Image by Helen M. Giesel.
}
\label{fig:QuIET_3d}
\end{figure}

The purpose of this article is to communicate the details of this proposal,
including several potential chemical structures to facilitate fabrication and
testing of this device. In Section \ref{theory}, we describe the theoretical
framework used to model the device. Section \ref{mechanism} explains the
QuIET's operating mechanism. Section \ref{implementations} discusses practical
implementations of the device. We conclude in Section \ref{conclusions}.

\section{Theoretical Model}
\label{theory}

The QuIET consists of a central molecular element, two leads chemically bonded
to the molecule, and a third lead, which can be coupled to the molecule either
capacitively or via tunneling. 
The Hamiltonian of this system can be written as the sum of three terms:
\begin{equation}
H=H_{mol}+H_{leads}+H_{tun}.
\end{equation}
The first is the 
$\pi$-electron molecular Hamiltonian
\begin{equation}
\label{Hm}
H_{mol}=\sum_{n\sigma}\varepsilon_n d_{n\sigma}^\dagger d_{n\sigma}
-\sum_{nm\sigma}\left(t_{nm}d_{n\sigma}^\dagger
  d_{m\sigma}+\mathrm{H.c.}\right)
+\sum_{nm}\frac{U_{nm}}{2}Q_nQ_m,
\end{equation}
where $d^\dagger_{n\sigma}$ creates an electron of spin $\sigma=\uparrow,\downarrow$ 
in the $\pi$-orbital of the $n$th carbon atom, and
$\varepsilon_n$ are the orbital energies.
We use a tight-binding model for the hopping
matrix elements with $t_{nm}=2.2\mbox{eV}$, $2.6\mbox{eV}$, or $2.4\mbox{eV}$ for 
orbitals
connected by a single bond, double bond, or within an aromatic ring,
respectively, and zero otherwise.
The final term of Eq.\ (\ref{Hm}) contains intra- and intersite
Coulomb interactions, as well as the electrostatic coupling to the
leads. The interaction energies are given by 
the Ohno parameterization \cite{ohno64,chandross97}:
\begin{equation}
\label{interactions}
U_{nm}=\frac{11.13\mathrm{eV}}{\sqrt{1+.6117\left(R_{nm}/\mathrm{\AA}\right)^2}},
\end{equation}
where $R_{nm}$ is the distance between orbitals $n$ and $m$.
\begin{equation}
Q_n=\sum_\sigma d_{n\sigma}^\dagger
d_{n\sigma}-\sum_\alpha C_{n\alpha}\mathcal{V}_\alpha/e -1
\end{equation}
 is an
effective charge operator \cite{stafford98} for orbital $n$, where
the second term represents a polarization charge.
Here $C_{n\alpha}$ is the capacitance between orbital $n$ 
and lead $\alpha$, chosen consistent 
with the interaction energies of Eq.\
(\ref{interactions}) and the geometry of the device, and $\mathcal{V}_\alpha$ is the voltage on lead
$\alpha$. $e$ is the magnitude of the electron charge.

Each metal lead $\alpha$ possesses a continuum of states, and their total Hamiltonian is
\begin{equation}
H_{leads}=\sum_{\alpha=1}^3\sum_{k\in\alpha\atop\sigma}\epsilon_k c_{k\sigma}^\dagger c_{k\sigma},
\end{equation}
where $\epsilon_k$ are the energies of the single-particle levels in the leads, 
and $c^\dagger_{k\sigma}$ is an electron creation operator.
Here leads 1 and 2 are the source and drain, respectively, and lead 3 is the control, or gate
electrode. 

Tunneling between molecule and leads is provided by the final term of the
Hamiltonian,
\begin{equation}
H_{tun}=\sum_{\langle n\alpha\rangle}\sum_{k\in\alpha\atop\sigma}\left(V_{nk}d_{n\sigma}^\dagger
  c_{k\sigma}+\mathrm{H.c.}\right),
\end{equation}
where $V_{nk}$ are the tunneling matrix elements from a level $k$ within
lead $\alpha$ to the nearby $\pi$-orbital $n$ of the molecule. Coupling of the leads to the molecule via
molecular chains, as may be desirable for fabrication purposes, can be
included in the effective $V_{nk}$, as can the effect of substituents (e.g., thiol groups)
used to bond the leads to the molecule \cite{tian98,nitzan01}. 

We use the non-equilibrium Green function
(NEGF) approach \cite{jauho94,datta95} to describe transport in this open quantum system.
The retarded Green function of the full system is
\begin{equation}
\label{Dyson}
G(E)=\left[E-H_{mol}-\Sigma(E)\right]^{-1},
\end{equation}
where $\Sigma$ is an operator, known as the retarded self-energy, describing the coupling of the molecule to 
the leads.
The QuIET is intended for use at room temperature, 
and operates in a voltage regime where there are no unpaired electrons in the molecule.
Thus lead-lead and lead-molecule correlations, such as the Kondo effect, do not play an important role.
Electron-electron interactions may therefore be included via the self-consistent Hartree-Fock
method. 
$H_{mol}$ is replaced by the corresponding 
mean-field Hartree-Fock
Hamiltonian $H_{mol}^{HF}$, which is quadratic in electron creation and annihilation operators,
and contains long-range hopping.  Within mean-field theory, the 
self-energy is a diagonal matrix
\begin{equation}
\label{self_energy}
\Sigma_{n\sigma,m\sigma'}(E)=\delta_{nm}\delta_{\sigma\sigma'}\!\sum_{\langle a\alpha\rangle} \delta_{na}
\Sigma_\alpha (E),
\end{equation}
with nonzero entries on the $\pi$-orbitals adjacent to each lead $\alpha$:
\begin{equation}
\label{self_energy_diag}
\Sigma_\alpha(E) =
\sum_{k \in \alpha \atop \langle n\alpha\rangle} \frac{|V_{nk}|^2}{E - \epsilon_k + i 0^+}.
\end{equation}

The imaginary parts of the self-energy matrix elements determine
the Fermi's Golden Rule tunneling widths
\begin{equation}
\Gamma_\alpha(E)\equiv -2\,\mbox{Im}\Sigma_\alpha(E)
=2\pi\sum_{k\in\alpha}|V_{nk}|^2\delta\left(E-\epsilon_{k}\right).
\end{equation}
As a consequence, the molecular density of states 
changes from a discrete spectrum of delta functions to a continuous, width-broadened distribution. 
We take the broad-band limit \cite{jauho94}, treating $\Gamma_\alpha$ as constants characterizing the coupling of the leads to the molecule. 
Typical estimates \cite{nitzan01} using the  method of Ref.\ \cite{mujica94} yield
$\Gamma_\alpha \lesssim 0.5\mbox{eV}$, but values as large as 1eV have been suggested \cite{tian98}. 

The effective hopping and orbital energies in 
$H_{mol}^{HF}$ depend on the equal-time correlation functions, which are found in the NEGF approach 
to be
\begin{equation}
\label{correlations}
\langle d_{n\sigma}^\dagger
d_{m\sigma}\rangle=\sum_{\langle a\alpha\rangle}\frac{\Gamma_\alpha}{2\pi} \int_{-\infty}^\infty \!\!\! dE \,
G_{n\sigma,a\sigma}(E)G^*_{a\sigma,m\sigma}(E)f_\alpha(E),
\end{equation}
where $f_\alpha(E)=\{1+\exp[(E-\mu_\alpha)/k_B T]\}^{-1}$ is the Fermi function for lead $\alpha$. 
Finally, the Green function is determined 
by iterating the self-consistent loop, Eqs.\ (\ref{Dyson})--(\ref{correlations}).

The current in lead $\alpha$ is given by the multi-terminal current formula \cite{buttiker86}
\begin{equation}
\label{l-b}
I_\alpha=\frac{2e}{h}\sum_{\beta=1}^3\int_{-\infty}^\infty
dE\;T_{\beta\alpha}(E)\left[f_\beta(E)-f_\alpha(E)\right],
\end{equation}
where 
\begin{equation}
T_{\beta\alpha}(E)=\Gamma_\beta\Gamma_\alpha|G_{ba}(E)|^2
\end{equation}
is the transmission probability \cite{datta95} from lead $\alpha$ to lead $\beta$, 
and $a\, (b)$ is the orbital coupled to lead $\alpha\, (\beta)$.
Similar mean-field NEGF calculations have been widely used to treat two-terminal transport through single molecules \cite{nitzan03}.

\section{Switching Mechanism}
\label{mechanism}

\begin{figure}[t]
  \setlength\unitlength{.333\columnwidth}
  \begin{picture}(3,3)
    \put(0,2){\includegraphics[width=\columnwidth,height=\unitlength]{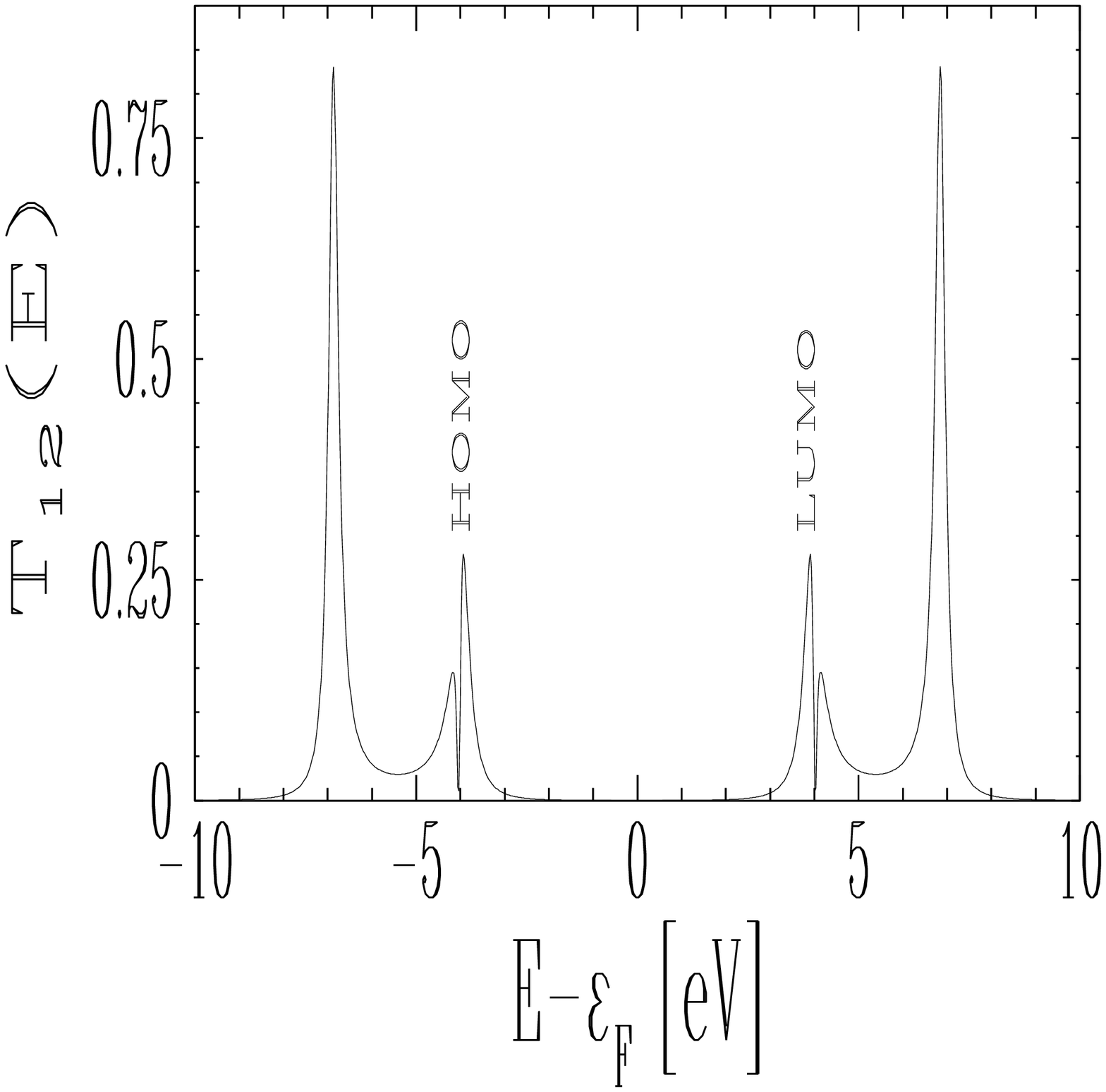}}
    \put(0,1){\includegraphics[width=\columnwidth,height=\unitlength]{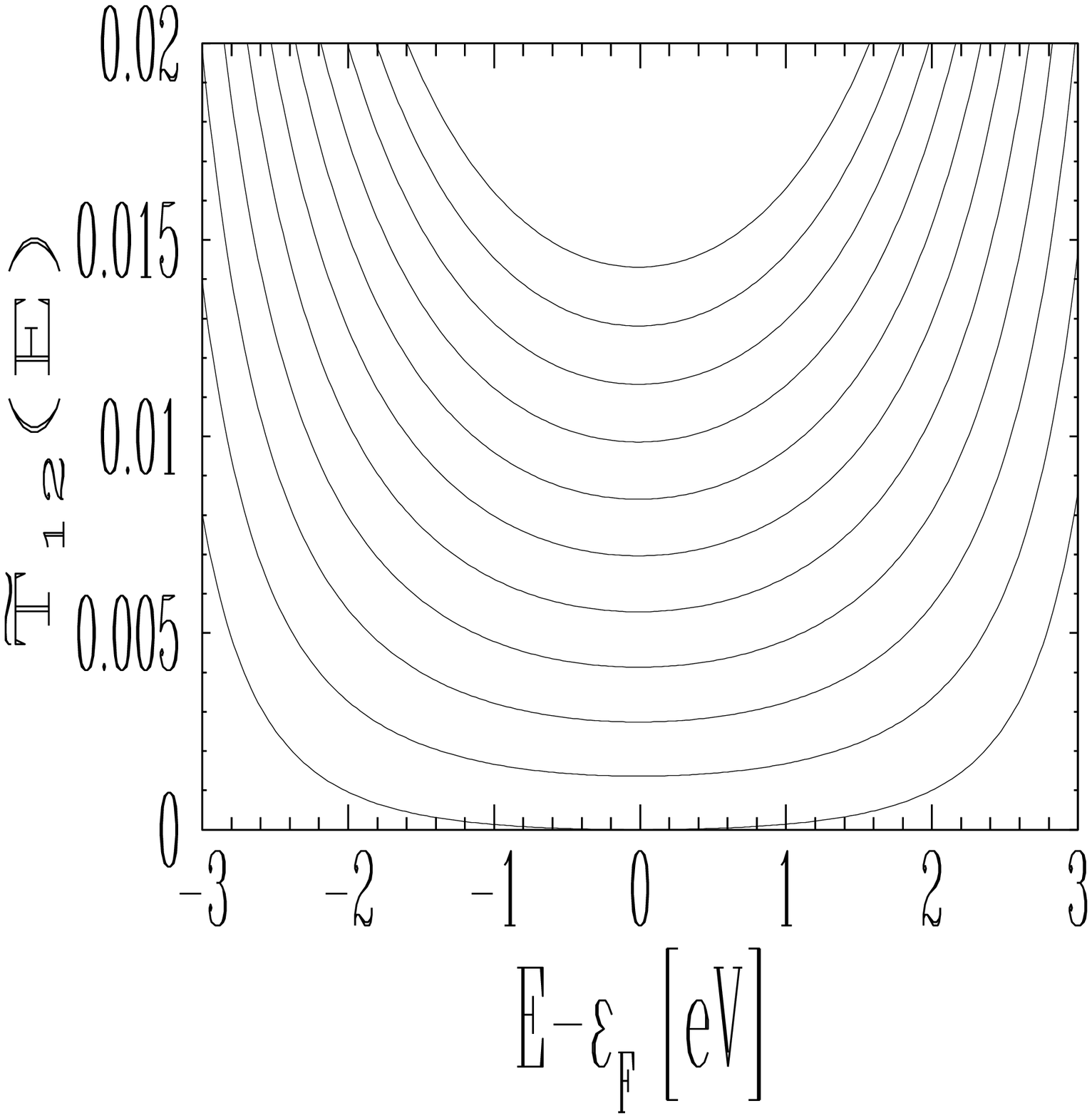}}
    \put(0,0){\includegraphics[width=\columnwidth,height=\unitlength]{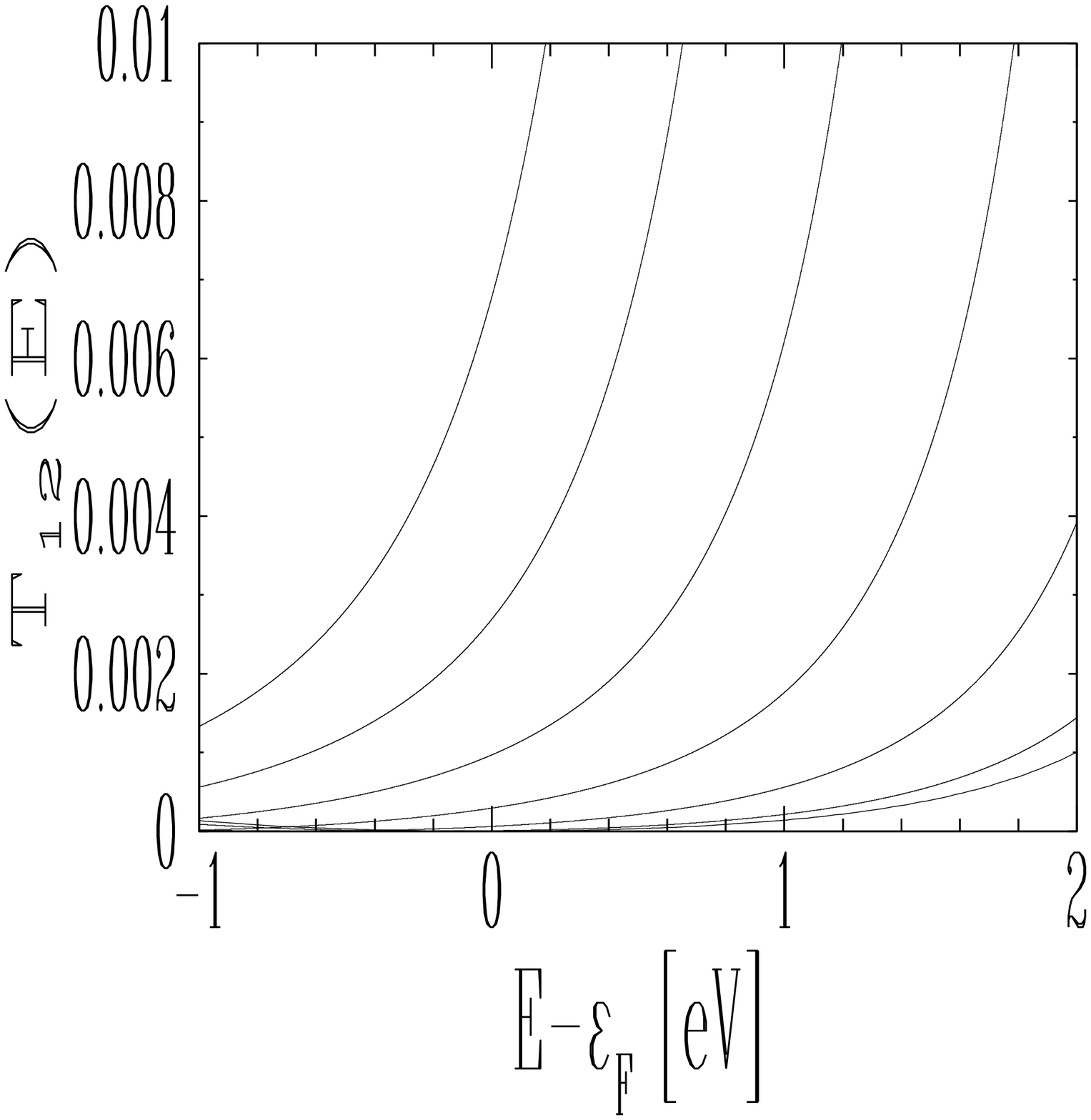}}
    \put(0,2.95){(a)}
    \put(0,1.95){(b)}
    \put(0,.95){(c)}
    \put(.55,.45){\includegraphics[width=.16\columnwidth,keepaspectratio=true]{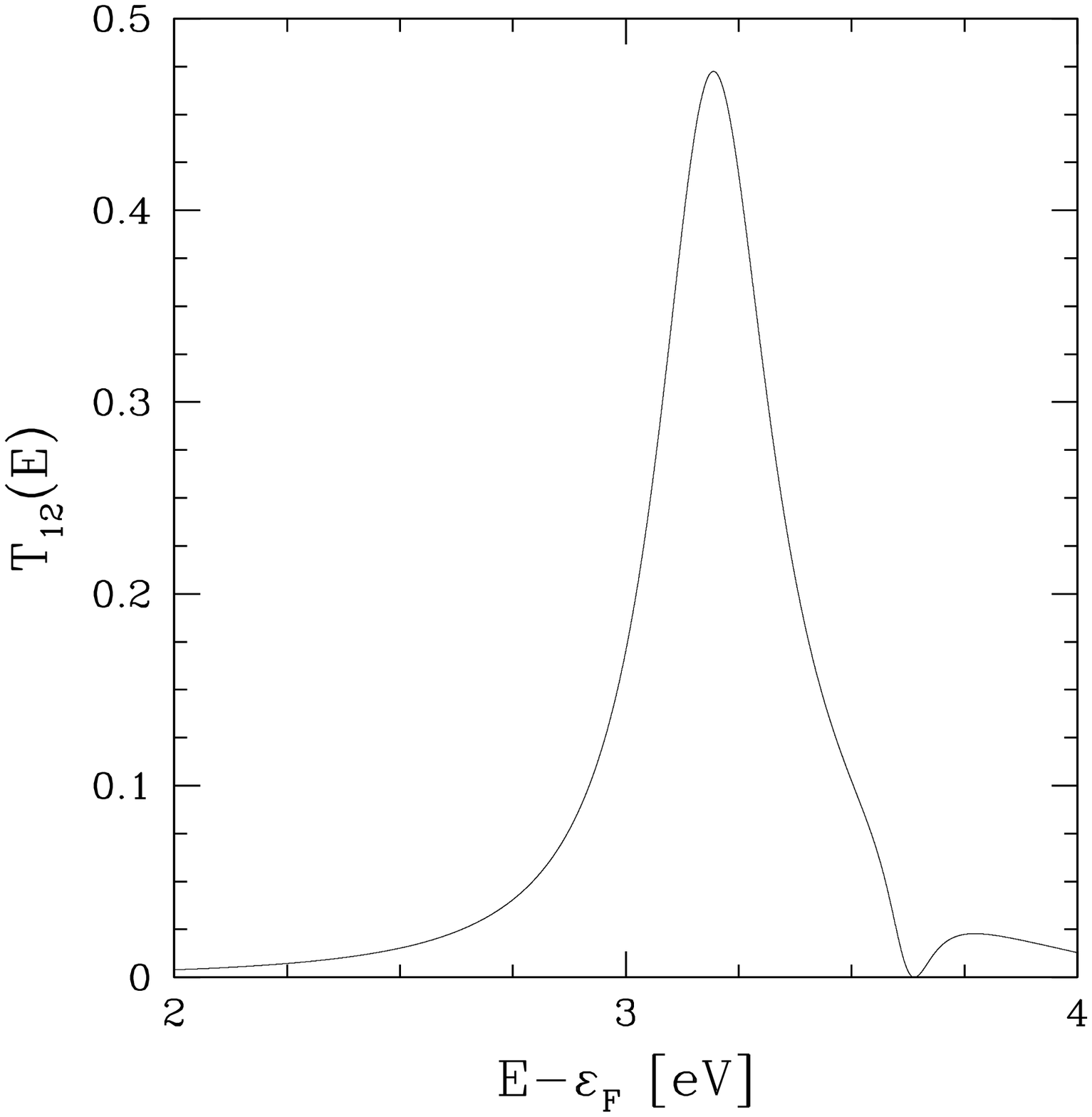}}
  \end{picture}
  \caption{Effective transmission probability $\tilde{T}_{12}$
  of the device shown in Fig.~\ref{fig:QuIET_3d}, at room temperature,
  with $\Gamma_1=1.2\mathrm{eV}$ and
  $\Gamma_2=.48\mathrm{eV}$.
  Here $\varepsilon_F$ is the Fermi level of the molecule. (a) $\Sigma_3=0$; 
  (b) $\Sigma_3=-i\Gamma_3/2$, where
  $\Gamma_3=0$ in the lowest curve, and increases by .24eV in each
  successive one; 
 (c) $\Sigma_3$ is given by Eq.\ (\ref{realsigma}) with a single resonance at
 $\varepsilon_\nu=\varepsilon_F + 4\mbox{eV}$. Here $t_\nu=0$ in the lowest curve, and increases by 0.5eV in each 
successive curve. Inset:  Full vertical scale for $t_\nu=1\mathrm{eV}$.
From Ref.\ \cite{QuIETNL}.
}
  \label{suppression}
\end{figure}

The QuIET exploits quantum interference stemming from the symmetry of monocyclic aromatic annulenes
such as benzene. 
Quantum transport through single benzene molecules with two metallic leads connected at {\it para} positions has
been the subject of extensive experimental and theoretical investigation \cite{nitzan03}; however, a QuIET based on 
benzene requires the source (1) and drain (2) to be connected at {\it meta} positions, as illustrated in Fig.~\ref{fig:QuIET_3d}.
The transmission probability $T_{12}$ of this device, for $\Sigma_3=0$, 
is shown in Fig.\ \ref{suppression}.
Due to the molecular symmetry \cite{ratner02}, there is a node in $T_{12}(E)$, located midway between the HOMO and LUMO energy
levels (see Figure \ref{suppression}b, lowest curve).   This mid-gap node, at the Fermi level of the molecule,
plays an essential role in the operation of the QuIET.

The existence of a transmission node for the meta connection can be understood in terms of
the Feynman path integral formulation of quantum mechanics \cite{feynman}, according to which
an electron 
moving from lead 1 to lead 2
takes
all possible paths within the molecule; observables relate only to the complex
sum over 
paths.
In the absence of a third lead ($\Sigma_3=0$), these paths all lie within the benzene ring.
An electron 
entering
the molecule at the Fermi level has de Broglie wavevector
$k_F=\pi/2d$, where $d=1.397\mathrm{\AA}$ is the intersite spacing of
benzene (note that $k_F$ is a purely geometrical quantity, which is unaltered by electron-electron interactions \cite{luttinger}).
The two most direct paths through the ring have lengths $2d$ and $4d$, with a phase 
difference $k_F 2d = \pi$, so they interfere destructively.
Similarly, all of the paths through the ring 
cancel exactly in a pairwise fashion, leading to a node in the transmission probability at $E=\varepsilon_F$. 

This transmission node can be lifted 
by introducing decoherence or elastic scattering that break the molecular symmetry. 
Figures \ref{suppression}b and c illustrate the effect of coupling a third lead to the molecule,
introducing a complex self-energy $\Sigma_3(E)$ on the $\pi$-orbital adjacent to that connected to lead 1 or 2.  
An imaginary self-energy $\Sigma_3=-i\Gamma_3/2$ corresponds to coupling a third metallic lead directly to the
benzene molecule, as shown in Fig.~\ref{fig:QuIET_3d}. 
If the third lead functions as an infinite-impedance voltage probe, the effective two-terminal transmission is \cite{buttiker88}
\begin{equation}
\tilde{T}_{12}=T_{12}+\frac{T_{13}T_{32}}{T_{13}+T_{32}}.
\end{equation}
The third lead introduces decoherence \cite{buttiker88} and additional paths that are not canceled, thus allowing current to flow,
as shown in Fig.~\ref{suppression}b.
As a proof of principle, a QuIET could be constructed using a scanning
tunneling microscope tip as the third lead (cf.\ Figure \ref{fig:QuIET_3d}), with tunneling coupling $\Gamma_3(x)$
to the appropriate $\pi$-orbital of the benzene ring, 
the control variable $x$ being the piezo-voltage controlling the tip-molecule distance.

By contrast,
a real self-energy $\Sigma_3$ introduces elastic scattering, which can also break the molecular symmetry.
This can be achieved by attaching a second molecule to the benzene ring, for example an alkene chain. 
The retarded self-energy due to the presence of a second molecule is 
\begin{equation}
\label{realsigma}
\Sigma_3(E)=\sum_\nu\frac{|t_\nu|^2}{E-\varepsilon_\nu+i0^+},
\end{equation}
where $\varepsilon_\nu$ is the energy of the $\nu$th molecular orbital of the second molecule, and $t_\nu$ is the hopping integral
coupling this orbital with the neighboring $\pi$-orbital of the benzene ring.
Figure \ref{suppression}c shows the transmission probability $T_{12}(E)$ in the vicinity of the Fermi energy of the molecule, for
the case of a single side-orbital at $\varepsilon_\nu=\varepsilon_F+4\mbox{eV}$.  As the coupling $t_\nu$ is increased, the node 
in transmission at $E=\varepsilon_F$ is lifted due to scattering from the side orbital.
The sidegroup introduces Fano antiresonances \cite{sols89,clerk01}, 
which suppress
current through one arm of the annulene, thus lifting the
destructive interference.  Put another way, the second molecule's orbitals
hybridize with those of the annulene, and a state that connects leads 1 and 2 is created in the gap [see Figure \ref{suppression}c (inset)]. 
In practice, either $t_\nu$ or $\varepsilon_\nu$ might be varied to control the
strength of Fano scattering.

Tunable current suppression occurs over a broad energy range, 
as shown in Fig.~\ref{suppression}b; 
the QuIET functions with any metallic leads 
whose work function lies within the annulene gap.
Fortunately, this is the case for many bulk metals, among them palladium,
iridium, platinum, and gold \cite{marder00}. 
Appropriately doped semiconductor electrodes \cite{piva05} could also be used.

\begin{figure}
\includegraphics[width=\columnwidth,height=.5\columnwidth]{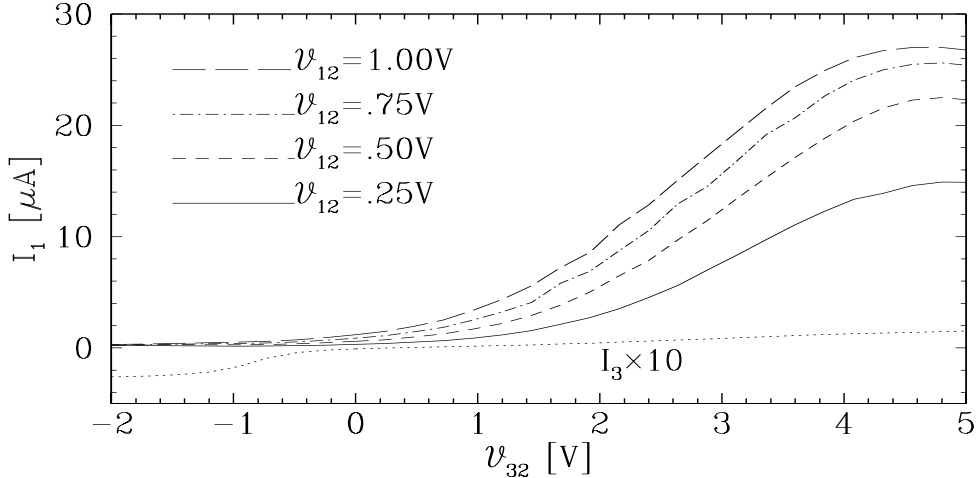}
\caption{Room temperature $I$--$\mathcal{V}$ characteristic of a QuIET based on sulfonated vinylbenzene.
  The current in lead 1 is shown, where
  $\mathcal{V}_{\alpha\beta}=\mathcal{V}_\alpha-\mathcal{V}_\beta$. Here,
  $\Gamma_1=\Gamma_2=$1eV. 
  $\Gamma_3$ is taken as .0024eV, which allows a small
  current in the third lead, so that the device amplifies current.
  A field-effect device with almost identical $I$--$\mathcal{V}$ can be achieved by taking
  $\Gamma_3=0$. The curve for $I_3$ is for the case of 1.00V bias voltage;
  $I_3$ for other biases look similar.
  From Ref.\ \cite{QuIETNL}.
}
\label{IV}
\end{figure}

We show in Fig.~\ref{IV} the $I$--$\mathcal{V}$ characteristic of a QuIET based on sulfonated vinylbenzene.
The three metallic electrodes were taken as bulk gold, with
$\Gamma_1=\Gamma_2=1\mbox{eV}$, while $\Gamma_3=.0024\mbox{eV}$, so that the coupling
of the third electrode to the alkene sidegroup is primarily electrostatic.
The device characteristic resembles that of a macroscopic transistor.
As the voltage on lead 3 is increased, scattering from the antibonding orbital of the alkene sidegroup increases as it approaches
the Fermi energies of leads 1 and 2,
leading to a broad peak in the current.
For $\Gamma_{1,2} \gg \Gamma_3\neq 0$, the device amplifies the current
in the third lead (dotted curve), emulating a bipolar junction
transistor.
Alkene chains containing 4 and 6 carbon atoms were also studied, 
yielding devices with characteristics similar to that shown in Fig.~\ref{IV}, with the maximum current $I_1$ shifting to smaller
values of $\mathcal{V}_{32}$ with increasing chain length.  
As evidence that the transistor behavior shown in Fig.~\ref{IV} is due to the tunable interference
mechanism discussed above, we point out that if hopping between the benzene ring
and the alkene sidegroup is set to zero, so that the coupling of the sidegroup to benzene is purely electrostatic,
almost no current flows between leads 1 and 2.

\begin{figure}
\includegraphics[width=\columnwidth,height=.5\columnwidth]{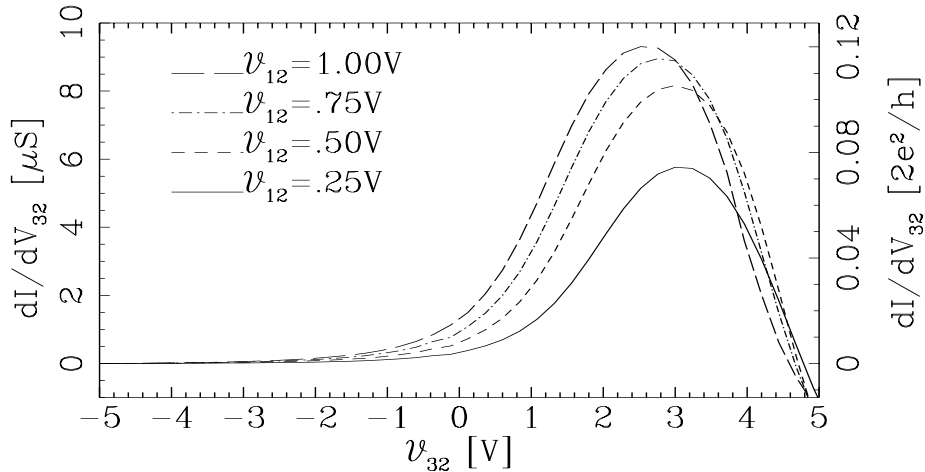}
\caption{Transconductance ${dI}/{d\mathcal{V}_{32}}$ of a QuIET based on sulfonated vinylbenzene
  with $\Gamma_3=0$. The characteristic is similar to that of a
  field-effect transistor, \emph{i.e.} $I_3=0$ while $I_1=-I_2=I$. As in
  Fig.\ \ref{IV}, $\Gamma_1=\Gamma_2=1\mathrm{eV}$, and the calculation was
  done for room temperature. 
From Ref.\ \cite{QuIETNL}.
}
\label{fig:transconductance}
\end{figure}

For $\Gamma_3=0$, $I_3=0$ and the QuIET behaves as a field-effect
transistor. The transconductance ${dI}/{d\mathcal{V}_{32}}$ of such a device
is shown in Fig.~\ref{fig:transconductance}. For comparison,
we note that an ideal single-electron transistor \cite{SET:ref} with $\Gamma_1=\Gamma_2=1e\mathrm{V}$
has peak transconductance $(1/17)G_0$ at bias voltage
$.25\mathrm{V}$, and $(1/2)G_0$ at bias $1\mathrm{V}$, where $G_0=2e^2/h$ is
the conductance quantum. For low biases, the proposed QuIET thus has a higher
transconductance than the prototypical 
nanoscale amplifier, while even for
large biases its peak transconductance is
comparable. Likewise, the load resistances required for a QuIET to have gain
(load times transconductance) greater than one while in its ``on'' state are
comparable to other nanoscale devices, $\sim 10/G_0$.

Operation of the QuIET does not depend sensitively 
on the magnitude of the lead-molecule coupling $\bar{\Gamma}=\Gamma_1\Gamma_2/(\Gamma_1+\Gamma_2)$.
The current through the device decreases with decreasing $\bar{\Gamma}$, but aside from that, the 
device characteristic was found to be qualitatively similar when $\bar{\Gamma}$ was varied over one order of magnitude.
The QuIET is also insensitive to molecular vibrations: only vibrational modes that simultaneously
alter the carbon-carbon bond lengths and break the six-fold symmetry within the benzene component can 
cause decoherence in a benzene `interferometer.'  Such modes are only excited at temperatures greater than
about 500K.

The position of the third lead affects the degree to
which destructive interference is suppressed.  For benzene, the most effective
location for the third lead is shown in Fig.~\ref{fig:QuIET_3d}. It may also be
placed at the site immediately between leads 1 and 2, but the transistor effect is somewhat
reduced, since coupling to the charge carriers is less.
The third, three-fold symmetric configuration of leads completely
decouples the third lead from electrons traveling between the first two leads.
For each monocyclic aromatic annulene, 
one three-fold symmetric lead configuration exists, yielding no transistor behavior.

\begin{figure}
\includegraphics[keepaspectratio=true,width=\columnwidth]{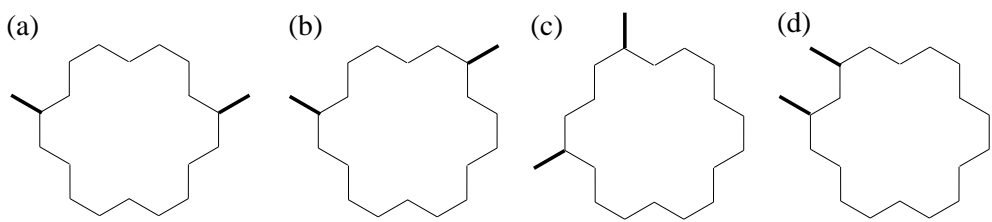}
\caption{Source-drain lead configurations possible in a QuIET
  based on [18]-annulene. The bold lines represent the positioning of the two
  leads. Each of the four arrangements has a different phase difference
  associated with it: (a) $\pi$; (b) $3\pi$; (c) $5\pi$; and (d) $7\pi$.
  From Ref.\ \cite{QuIETNL}.
}
\label{fig:18}
\end{figure}

The QuIET's operating mechanism, tunable coherent current suppression,
occurs over a broad energy range within the gap of each monocyclic aromatic annulene;
it is thus a {\em very robust effect, insensitive to moderate fluctuations of the electrical environment of the molecule}.
Although based on an entirely different, quantum mechanical, switching mechanism, the QuIET nonetheless reproduces
the functionality of macroscopic transistors on the scale of a single molecule.

\section{Implementations}
\label{implementations}

\begin{figure}
\includegraphics[keepaspectratio=true,width=\columnwidth]{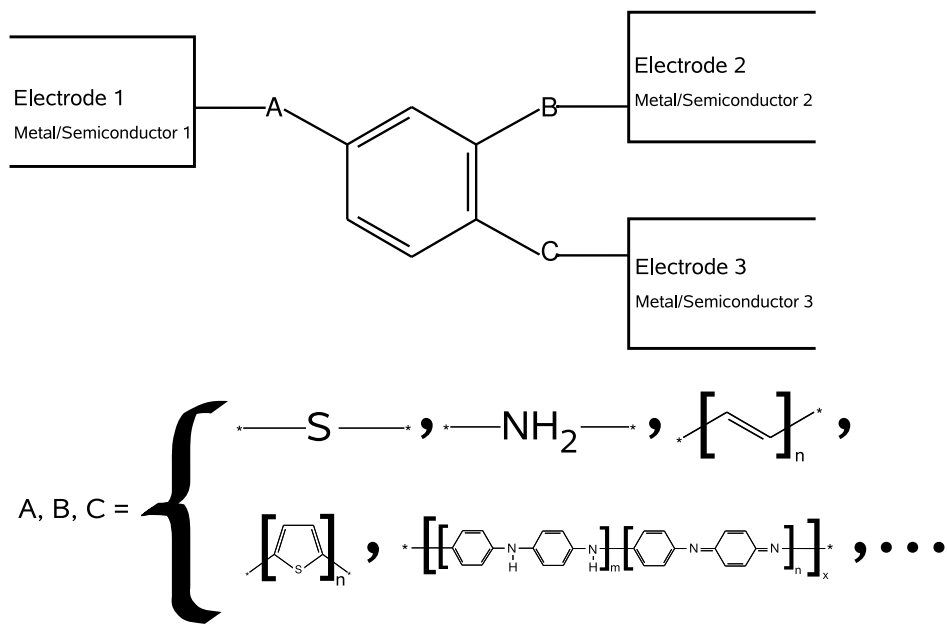}
\caption{Schematic of various QuIETs based on a benzene ring. A, B, and C
  represent the various substituents which may be placed in series between the
  ring and each lead. In particular, the conducting polymers like polyaniline
  and polythiophene may be useful in overcoming the ``third lead'' problem.}
\label{wires}
\end{figure}

As daunting as the fundamental problem of the switching mechanism is the
practical one of nanofabrication. 
The QuIET requires a third lead coupled locally to the central molecule, and,
while there has recently been significant progress in that direction \cite{piva05,wolkow06,grutter07}, to date,
only two-lead single molecular devices, sometimes with global gating, have
been achieved \cite{nitzan03}. With this in mind, we turn to potential practical realizations
of the device.

Using novel fabrication techniques, such as ultra-sharp STM tips
\cite{wolkow06} or substrate
pitting \cite{grutter07}, it may soon be possible to attach multiple leads to large
molecules. Fortunately, the QuIET mechanism applies not only to benzene, but to any monocyclic aromatic annulene with leads 1 and 2 positioned so the two most
direct paths have a phase difference of $\pi$. Furthermore,
larger molecules have other possible lead configurations, based on phase
differences of $3\pi$, $5\pi$, etc.; as an example, Figure \ref{fig:18} shows
the lead configurations for a QuIET based on [18]annulene. Other large
ring-like molecules, such as [14]annulene and divalent metal-phthalocyanine,
would also serve well.

\begin{figure}
\includegraphics[keepaspectratio=true,width=\columnwidth]{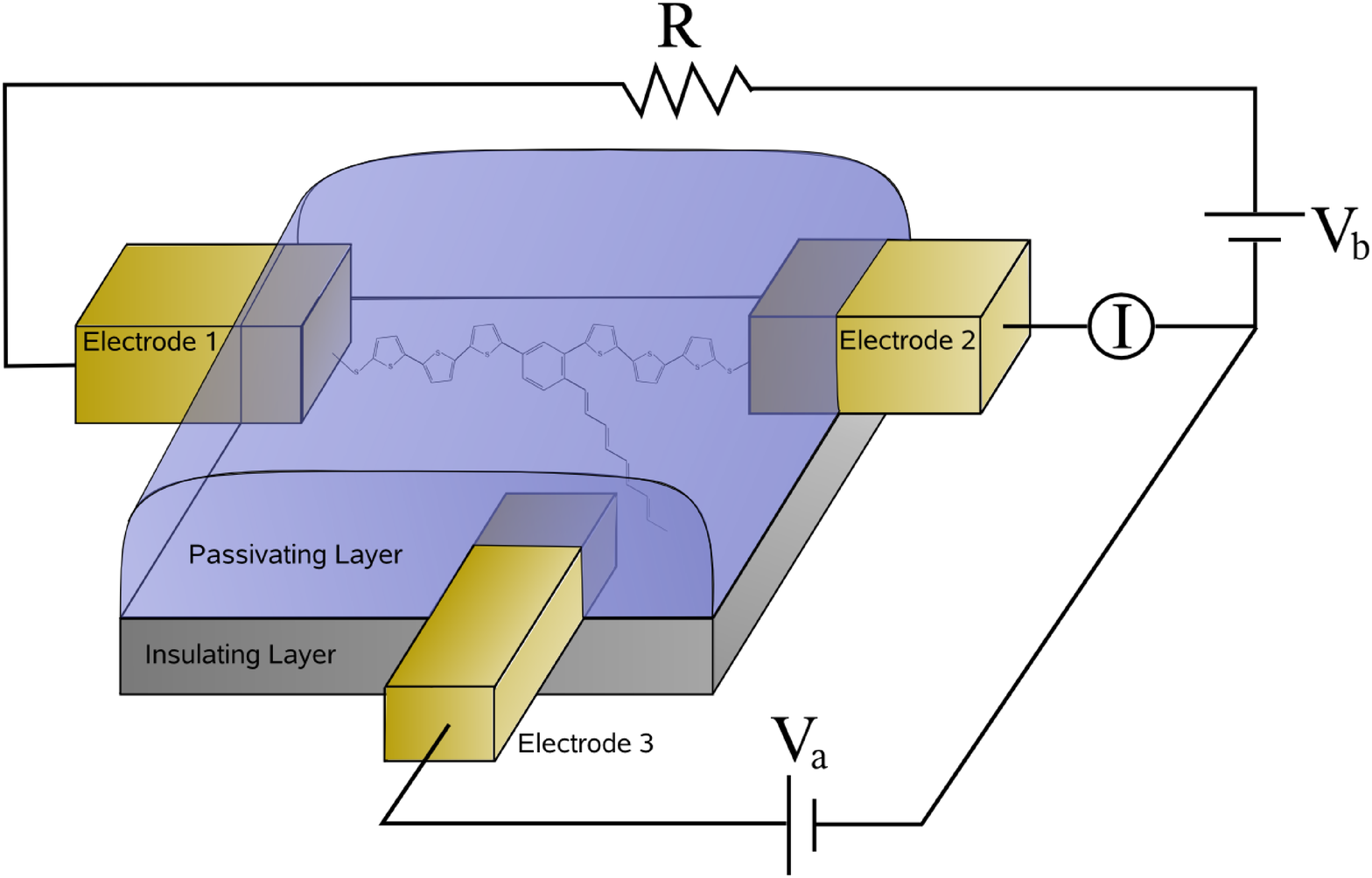}
\caption{
Possible embodiment of a QuIET integrated with conventional circuitry on a chip.
The source (1) and drain (2) electrodes are connected via conducting polymers (in this case, polythiophene)
to the central aromatic ring, while the gate electrode (3) is coupled electrostatically to an alkene sidegroup.
  }
\label{embodiment}
\end{figure}

Another method of increasing the effective size of the molecule is to introduce
molecular wires linking the central ring and leads (see Figures
\ref{wires} and \ref{embodiment}). Conducting polymers, such as polythiophene or polyaniline, are
ideal for this task. Such changes can be absorbed into the diagonal elements
of the self-energy $\Sigma(E)$, and so only modify $G(E)$ locally. As such,
while they can significantly modify the on-resonance behavior of a molecular
device, off-resonance function is largely unaltered. In particular, the
transmission node at the centre of the gap is unaffected.
An example
of such a QuIET integrated with conventional circuitry on a chip is shown
in Figure \ref{embodiment}.  

\section{Conclusions}
\label{conclusions}
The quantum interference effect transistor represents one way to
simultaneously overcome the problems of scalability and power dissipation which
face the next generation of transistors. Because of the exact symmetry
possible in molecular devices, it possesses a perfect mid-gap transmission node,
which serves as the off state for the device. Tunably introduced decoherence
or elastic scattering can lift this quantum interference effect, with the result
of current modulation. Furthermore, a vast variety of potential chemical
structures possess the requisite symmetry, easing fabrication difficulties. In
particular, molecular wires, such as conducting polymers, can be used to
extend the molecule to arbitrary size.

\ack
This work was supported in part by National Science Foundation Grant Nos.\ PHY0210750, DMR0312028, and DMR0406604.

\end{document}